\documentclass[12pt]{article}
\usepackage{amsmath,amssymb,graphicx}
\def\be{\begin{equation}}
\def\ee{\end{equation}}
\title{Addendum to ``Thin-shell wormholes supported by ordinary matter in Einstein--Gauss--Bonnet gravity''\footnote{arXiv:0710.2041v3[gr-qc] (Phys. Rev. D {\bf 76}, 087502 (2007);  D {\bf 77}, 089903(E) (2008).)}}
\author{Claudio Simeone\thanks{E-mail: csimeone@df.uba.ar. }\\
{\small  IFIBA-CONICET, and Departamento de F\'{\i}sica,}\\
{\small Facultad de Ciencias Exactas y 
Naturales, Universidad de Buenos Aires}\\ 
{\small Ciudad Universitaria, Pabell\' on I, 1428, 
Buenos Aires, Argentina}} 

\begin{document}
\maketitle

\begin{abstract}

\noindent Thin-shell wormholes are constructed starting from the exotic branch of Wiltshire spherically symmetric solution of Einstein--Gauss--Bonnet gravity. The energy-momentum tensor of the shell is studied, and it is shown that configurations supported by matter satisfying the energy conditions exist for certain values of the parameters. Differing from the previous result associated to the normal branch of Wiltshire solution, this is achieved for small positive values of the Gauss--Bonnet parameter and for vanishing charge.

\end{abstract}

\vskip0.7cm



\baselineskip.25in

\section{Introduction}

 The equations of gravitation admit topologically non trivial  solutions, known as Lorentzian wormholes, which connect two regions of the same universe (or of two universes) by a throat. In the case of compact wormholes, the throat is defined as a minimal area surface \cite{motho,book}. Wormholes could present features of remarkable interest, including the possibility of time travel (see Refs.  \cite{morris-novikov}); but there is an important objection against their actual existence: in Einstein gravity the  flare-out condition \cite{hovis1} to be  satisfied at the throat of compact wormholes requires  exotic matter, that is, matter violating the energy conditions \cite{book}. However, in the last few years it was shown  that in pure Gauss--Bonnet gravity  \cite{gra-wi} wormholes  could exist even with no matter (see also Refs. \cite{7,maeda}). Moreover, in our recent paper \cite{risi08} we found 5-dimensional Einstein--Maxwell--Gauss--Bonnet spherical thin-shell wormholes supported by ordinary matter, which were later proven  to be stable under perturbations preserving the symmetry \cite{maz}.  But our non exotic construction presented two non desirable aspects: it required a very large charge (about the critical value) and a negative Gauss--Bonnet parameter $\alpha$.  Here we thus revisit thin-shell wormholes in Einstein--Gauss--Bonnet gravity starting from a different branch of Wiltshire solution \cite{wilt}, and we focus in the character --exotic or normal-- of matter on the shell in relation with the parameters of the system. The Gauss--Bonnet terms of the equations of the theory are associated to the geometry, and not to an effective energy momentum tensor; thus we work with the  Darmois--Israel formalism  for thin shells \cite{3} generalized to include the Gauss--Bonnet contribution \cite{4}, which allows for a better understanding of the character of the matter on the shell. We  show that for certain values of the parameters,  thin-shell wormholes constructed from the exotic branch of Wiltshire solution can be supported by non exotic matter even in the case of  a relatively small and positive Gauss--Bonnet constant and even with a vanishing charge. We shall see that no exotic wormhole configurations are associated to a repulsive far behavior of the mass term of the exotic branch metric.

\section{Junction conditions of the theory}

 The Einstein--Gauss--Bonnet equations of gravitation in five spacetime dimensions have the form
\begin{equation}
R_{\mu\nu}-\frac{1}{2}g_{\mu\nu}R+ \Lambda g_{\mu\nu}+2\alpha H_{\mu\nu}=\kappa^{2}T_{\mu\nu}\label{field}
\end{equation}
where $\Lambda$ is the cosmological constant, $\kappa^2$ is proportional to the Newton constant (set equal to unity in the units used here) and $\alpha$ is the constant of dimension $(\rm{length})^2$ multiplying the Gauss--Bonnet terms 
\begin{equation} 
H_{\mu\nu}=RR_{\mu\nu}-2R_{\mu\alpha}R^{\alpha}_{\nu}-2R^{\alpha\beta}R_{\mu \alpha\nu\beta}+R_{\mu}^{\alpha\beta\gamma}R_{\nu\alpha\beta\gamma}-\frac{1}{4}g_{\mu\nu}(R^{2}-4R^{\alpha\beta}R_{\alpha\beta}+R^{\alpha\beta\gamma\delta}R_{\alpha\beta\gamma\delta}).
\end{equation}
The matching conditions relating the metrics at the two sides of a given surface $\Sigma$ with the character of matter on this surface (the energy-momentum tensor of the shell placed at $\Sigma$) are given by the  Darmois--Israel conditions generalized  to  Einstein--Gauss--Bonnet gravity. They were obtained in Ref. \cite {4} starting from the equations above and read
\begin{equation}
2\langle K_{ij}-K h_{ij}\rangle + 4\alpha \langle 3J_{ij}-Jh_{ij}+2P_{iklj}K^{kl}\rangle =-\kappa^{2}S_{ij},
\end{equation}
where  $\langle \cdot\rangle$ stands for  the jump of a given quantity across the  surface $\Sigma$ and $h_{ij}$ is the surface induced metric. The extrinsic curvature $K_{ij}$ is defined as \begin{equation}
{K}^{\pm}_{ij}=-n^{\pm}_{\alpha}\left(\frac{\partial^{2}X^{\alpha}}{\partial\xi^{i}\partial\xi^{j}}+\Gamma^{\alpha}_{\beta\gamma}\frac{\partial X^{\beta}}{\partial\xi^{i}}\frac{\partial X^{\gamma}}{\partial\xi^{j}}\right)_{r=a},
\end{equation}
where $n_\alpha$ are the components of the unit normals to  $\Sigma$, $\xi^i$ (latin indices) are the coordinates on this surface, and $X^\mu$ (greek indices) are the coordinates of the $(4+1)-$dimensional manifold. The divergence-free part of the Riemann tensor $P_{ijkl}$ and the tensor $J_{ij}$ are defined as follows:
\begin{equation}
P_{ijkl}=R_{ijkl}+(R_{jk}h_{li}- R_{jl}h_{ki})-(R_{ik}h_{lj}- R_{il}h_{kj})+\frac{1}{2}R(h_{ik}h_{lj}-h_{il}h_{kj}),
\end{equation}
\begin{equation}
J_{ij}=\frac{1}{3}\left[2KK_{ik}K^{k}_{j}+K_{kl}K^{kl}K_{ij}-2K_{ik}K^{kl}K_{lj}-K^{2}K_{ij}\right]
\end{equation}
(see \cite{4} and also \cite{mae}). 

The equations of Einstein--Gauss--Bonnet gravity admit spherically symmetric solutions of the form
\begin{equation}
{ds}^{2}=-f(r)\,dt^{2}+f^{-1}(r)\,dr^{2}+r^{2}d\Omega^{2}_{3}.\label{metric}
\end{equation}
A static wormhole throat can be defined by matching two equal exterior metrics of this form at a static spherical surface of  radius $a$ (see below). Starting from the expresions above, a quite long but straightforward calculation gives that the non-null components  $S_{i}^{j}$ of the energy-momentun tensor of the shell placed at $r=a$ are
\begin{equation}
S_\tau^\tau=-\sigma=\frac{\sqrt{f(a)}}{8\pi a}\left[ 6-\frac{4\alpha}{a^2}\left(2f(a)-6\right)\right],\label{s0}
\end{equation}
\begin{equation}
S_\theta^\theta=S_\chi^\chi=S_\varphi^\varphi=p=\frac{\sqrt{f(a)}}{8\pi a}\left[ 4+a\frac{f^{'}(a)}{f(a)}-\frac{4\alpha f^{'}(a)}{a}\left(\frac{f(a)-1}{f(a)} \right)\right],\label{p0}
\end{equation}
where a prime denotes a radial derivative.  Note that in the Einstein's gravity framework ($\alpha=0$), even in $4+1$ dimensions, the energy density could never be positive, thus forbidding wormholes supported by non exotic matter. However, the second term within the bracket of Eq. (\ref{s0}) changes this drastically.

\section{Wormhole construction and characterization}

The metric function $f(r)$ of the general solution of Einstein--Maxwell--Gauss--Bonnet theory is given by \cite{wilt}
\begin{equation}
f(r)=1+\frac{r^{2}}{4\alpha}\left[1\mp\sqrt{1+\frac{16M\alpha}{\pi r^{4}}-\frac{8Q^{2}\alpha}{3 r^{6}}+\frac{4 \Lambda \alpha}{3}}\right],\label{f}
\end{equation}
where $Q$ the is charge. In the case of the $(-)$ sign in Eq. (\ref{f}), for both vanishing $Q$ and $\Lambda$  we obtain the far behavior
\be
f(r)\simeq 1-\frac{2M}{\pi r^2}
\ee
which allows to identify $M$ as the ADM mass. Instead, the metric corresponding to the $(+)$ sign in (\ref{f}) is usually denoted as the exotic branch, because for $Q=0$ and $\Lambda=0$ its far behavior is
\be
f(r)\simeq 1+\frac{2M}{\pi r^2}+\frac{r^2}{2\alpha}.
\ee
Thus for this solution the energy  $M$ enters in the geometry with the ``wrong'' sign; consequently, the mass term gives an attractive force on rest particles if $M<0$, while for $M>0$ the force is repulsive. Besides, this solution would not have the correct general relativity limit for $\alpha\to 0$; instead, it acquires an effective cosmological constant proportional to $\alpha^{-1}$. Note that for $M>0$ and $\alpha>0$ this geometry has no horizons, but it presents a naked singularity at the origin. In Ref. \cite{risi08} thin-shell wormholes were constructed starting from the $(-)$ branch of the metric; though they could be supported by non exotic matter, this was achieved at the price of a negative value of $\alpha$ and a very large charge. Therefore, here we shall explore the $(+)$ branch.

 To construct a thin-shell wormhole we proceed as usual (see, for example, the related work \cite{grg06} and references therein): We take two copies of the  spacetime given by (\ref{metric}) and (\ref{f})  with $\Lambda =0$ and remove from each manifold the five-dimensional regions corres\-ponding to  $r_{1,2}\leq a$. The resulting manifolds have boundaries given by the timelike hypersurfaces 
\begin{equation}
\Sigma_{1,2}=\left\{X/r_{1,2} = a\right\}.
\end{equation}
 Then the identification of these two surfaces gives a geodesically complete new mani\-fold $\cal M$ with a matter shell at the radius $r=a$, where the throat of the wormhole is located. Note that the throat does not connect two asymptotically flat regions\footnote{Thus some authors would object calling the configuration a true wormhole, but would speak about a geometry with a throat; see for instance Ref. \cite{star}.}, because the metric diverges for $r\to\infty$. But there are no singularities at finite distance, because the origin $r=0$ is removed in the mathematical construction. 

 We are interested in the possibility of matter fulfilling the weak energy condition \cite{book} $\sigma\geq 0$, $\sigma+p_i\geq 0$ (non exotic matter), which in our case reduces to $\sigma\geq 0$, $\sigma+p\geq 0$. From the equations (\ref{s0}) and (\ref{p0}) we obtain
\be
\sigma (a)+p(a)=-a\sigma'(a)/3,
\ee 
so that the energy condition is fulfilled if  $\sigma(a)\geq 0$ and simultaneously $\sigma'(a)\leq 0$. Replacing the metric for a vanishing cosmological constant in the expressions for the surface energy-momentum tensor, we obtain the condition for a positive energy density:
\begin{equation}
-8\alpha-2a^2+a^2\sqrt{1+\frac{16M\alpha}{\pi a^4}-\frac{8Q^2\alpha}{3a^6}}\geq 0.\label{s+}
\end{equation} 
We find that working with the exotic branch, for $\alpha>0$ the possibilities of $\sigma\geq 0$ are worse for $Q\neq 0$ than for $Q=0$.   Consequently, in the subsequent analysis we shall consider a null charge. Also note that if we impose $M<0$ in order to have a far behavior of the mass term corresponding to an attractive force on rest particles, then the condition $\sigma\geq 0$ cannot be fulfilled.  

 From Eq. (\ref{s+}) we can see that $\sigma>0$ imposes the necessary condition 
\begin{equation}
16M\alpha/(\pi a^4)>1.
\end{equation}
However, a large $\alpha$ drags the energy density to negative values because of the first term $-8\alpha$ of Eq. (\ref{s+}); then one expects at most a narrow interval of the parameters to be  compatible with a positive energy density. Besides, the sum $\sigma+p$ must also be positive to avoid exotic matter. From these considerations, it turns out that the most natural approach to answer the question about the existence of wormhole configurations supported by non exotic shells is to draw $\sigma (a)$ and $-a\sigma'(a)/3$ for different values of the parameters. We fix the scale with the mass and take $M=1$, and then consider different values of the Gauss--Bonnet constant $\alpha$. The results are illustrated in Figs. 1 and 2, where the energy density and the sum of it plus the pressure are displayed as functions of the throat radius  for some values of interest. In the mathematical cut and paste construction the singularity is removed, while the absence of horizons allows for the choice of  a throat radius as small as desired. We can see that for relatively small positive values of $\alpha$ the weak energy condition can be fulfilled, so that wormholes can be supported by non exotic matter. Also, we find that a smaller value of  $\alpha$ reduces the range of throat radii such that $\sigma\geq 0$ and $\sigma+p\geq 0$. As pointed above, for $M<0$ there is no solution for $\sigma\geq 0$ if we take $\alpha>0$, so that the energy condition could not be satisfied. Thus if the geometry has a far behavior consistent with an attractive force associated to the mass term, the shells supporting the wormholes must be exotic. Instead, if the far behavior corresponds to a repulsive force associated to the mass term, non exotic shells supporting the wormholes are possible. We remark that this result is achieved with a relatively small (compared with the mass) positive values of the Gauss--Bonnet parameter (see the values in the figures), and for a vanishing charge. 

\section{Summary}

We have been able to mathematically construct wormholes supported by a thin layer of normal matter starting from the exotic branch of the spherically symmetric solution of Einstein--Gauss--Bonnet gravity.  We have found that a far behavior of the mass term of the metric corresponding to a repulsion on rest particles is the only possibility compatible with a non exotic thin-shell wormhole. Besides, while in the case of the normal branch of the Wiltshire metric a negative $\alpha$ and a large charge were necessary to make an ordinary matter shell compatible with the existence of a throat \cite{risi08}, here we have shown that when the construction starts from the exotic branch the charge does not help but, instead, makes more difficult to fulfill the energy conditions, and that they can be satisfied with a positive Gauss--Bonnet parameter.

\begin{figure}[t!]
\centering
\includegraphics[width=10cm]{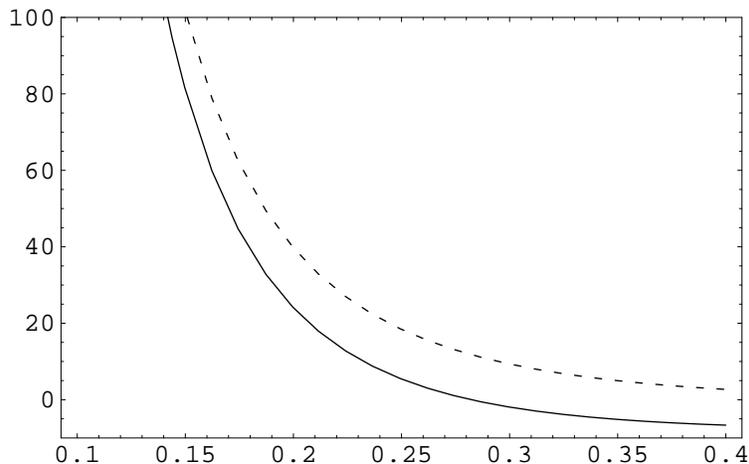}
\caption{Energy conditions at the throat: The solid line shows the energy density $\sigma$, and the dashed line shows the sum $\sigma+p$, as functions of the throat radius. The values of the parameters are $M=1$ and $\alpha=0.01$.}
\end{figure}

 \begin{figure}[t!]
\centering
\includegraphics[width=10cm]{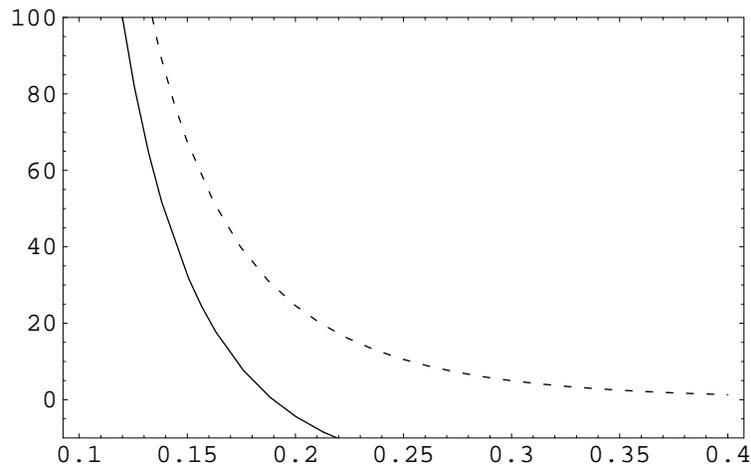}
\caption{Energy conditions at the throat: The solid line shows the energy density $\sigma$, and the dashed  line shows the sum $\sigma+p$, as functions of the throat radius.  The values of the parameters are  $M=1$  and $\alpha=0.001$.}
\end{figure}

\section*{Acknowledgements} I wish to thank E. F. Eiroa and G. Giribet for helpful discussions and comments. This work was supported by UBA and CONICET (Argentina).

\end{document}